\documentclass[12pt]{iopart}

\usepackage{amssymb}
\usepackage{graphicx}
\usepackage{color}
\usepackage[dvipsnames]{xcolor}
\usepackage{url}
\usepackage{hyperref}
\usepackage{color}
\usepackage[normalem]{ulem}
\usepackage{soul}

\newcommand{\exciting}{\texttt{exciting}}
\newcommand{\kpt}{\textbf{k}-point}

\newcommand{\kgrid}{\textbf{k}-grid}
\newcommand{\rgkmax}{\texttt{rgkmax}}

\newcommand{\eg}{{\it e.g.}, }

\begin{document}
	\submitto{Electronic Structure}

	\title{Ehrenfest dynamics implemented in the all-electron package \exciting}

	\author{Ronaldo Rodrigues Pela$^{1,2,3}$, Claudia Draxl$^{1,2}$}

	\address{$^1$ Physics Department and IRIS Adlershof, Humboldt-Universität zu Berlin, Zum Gro\ss en Windkanal 2, 12489 Berlin}
	\address{$^2$ European Theoretical Spectroscopy Facility (ETSF)}
	\address{$^3$ Supercomputing Department, Zuse Institute Berlin, Berlin, Germany.}
	\ead{ronaldo.rodrigues@zib.de}

	\begin{abstract}
		Ehrenfest Dynamics combined with real-time time-dependent density functional theory has proven to be a reliable tool to study non-adiabatic molecular dynamics with a reasonable computational cost. Among other possibilities, it allows for assessing in real time electronic excitations generated by ultra-fast laser pulses, as \eg\ in pump-probe spectroscopy, and their coupling to the nuclear vibrations even beyond the linear regime. In this work, we present its implementation in the all-electron full-potential package \exciting. Three cases are presented as examples: diamond and cubic BN relaxed after an initial lattice distortion, and cubic BN exposed to a laser pulse. Comparison with the Octopus code exhibits good agreement.
	\end{abstract}

	\maketitle

	\section{Introduction}
	Molecular dynamics (MD) is one of the most important techniques of computational materials science. In one of the first reports on a computer MD, Alder and Wainwright described it as tracking the movement of atoms which interact by means of a given potential \cite{Alder_1959,Andrade_2009}. In non-empirical MD, the potential is not modeled \emph{a priori} via some parametrized expression, but rather obtained ``on the fly'' through \emph{ab initio} calculations \cite{Andrade_2009}. Quite often then, the Born-Oppenheimer approximation (BOA) is employed. Thereby, each nucleus is treated as a point charge moving under the Coulomb potential due to the electrons and other nuclei, whereas the electrons are treated quantum mechanically with a wavefunction that responds adiabatically to the nuclei movement, remaining in an eigenstate in each time \cite{Scherrer_2017}.

	Despite many successful theoretical predictions, BOA is not appropriate for several applications, \eg those involving the interaction between electromagnetic waves and matter. Processes in which the excited electrons respond to a rapidly varying time-dependent field and then lose their energy by coupling to lattice vibrations, are non-adiabatic and thus have to be treated beyond the BOA \cite{Nelson_2020, Curchod_2013, Otero_2018,Pradhan_2018, You_2021}. Examples of such phenomena of growing interest are pump-probe experiments, in which an initial laser pulse excites the system, while a second one monitors its time-dependent relaxation towards the ground state (or a stable excited-state configuration).

    An efficient method to describe such non-adiabatic processes is the Ehrenfest Dynamics \cite{Ojanpera_2012, Wang_2011} that combines a classical treatment of the nuclei with a non-adiabatic evolution of electronic wavefunctions -- usually carried out in the framework of real-time time-dependent density-functional theory (RT-TDDFT). This approach has been successfully applied in many recent studies, as diverse as electron injection into BN nanotubes with thousands of atoms \cite{Andermatt_2016}, collisions between oxygen and graphite clusters \cite{Isborn_2007}, excited carrier dynamics in carbon nanotubes \cite{Miyamoto_2006}, ion collisions with carbon nanostructures \cite{Krasheninnikov_2007}, photoinduced dynamics in semiconductor quantum dots \cite{Prezhdo_2009}, ultrafast electron injection from a PbSe quantum dot into a TiO$_2$ surface \cite{Long_2011}, electron and hole transfer at a polymer-carbon nanotube heterojunction \cite{Long_2014}, or chlorine ion colliding to a graphene nanoflake consisting of hundreds of atoms \cite{Weile_2018}.

	A non-negligible aspect in computational materials science, in particular considering that we have entered the big-data era, is the validation of data. In electronic- structure theory, there are different approaches to solve the Kohn-Sham (KS) equations of DFT. Linearized augmented planewaves (LAPW) combined with local-orbitals (lo) can be regarded as the basis to describe KS wavefunctions with ultimate precision, thus serving as a reference for other implementations or approximations \cite{Lejaeghereaad_2016,Gulans_2018}. In this sense, the full-potential LAPW+lo \exciting{} code \cite{Gulans_2014} can be considered as reference for high-precision calculations. Moreover, it is open source and user-friendly, with more than 50 tutorials  explaining its features and how to explore them \cite{exciting_webpage}. In this manuscript, we report how we combine Ehrenfest Dynamics with RT-TDDFT, another most recent implementation \cite{Pela_2021}. To this extent, we briefly summarize the theory and provide the formalism behind our implementation together with LAPW-specific derivations. Finally, we show three different examples of applications, before drawing our conclusions.

	\section{Theoretical framework}\label{sec-theory}
	Within Ehrenfest dynamics, the nuclei behave as classical repelling point charges that move being submitted to an external electric field and the one imposed by themselves. The electrons dynamics, in turn, is typically modeled by RT-TDDFT. Considering the external electric field given in the gauge $\mathbf{E}(t) = -\frac{1}{c}\frac{d \mathbf{A}(t)}{dt}$,  $\mathbf{A}(t)$ being the vector potential and $c$ the speed of light, then the action integral for the total system of nuclei and electrons can be written as
	\cite{Saalmann_1996,Kunert_2003}
	\begin{eqnarray}\label{eq-action}
	\mathcal{A} &=& \int_{t_1}^{t_2}\mathrm{d}t \sum_J \left[ \frac{M_J\dot{\mathbf{R}}_J^2}{2} + \frac{Z_J}{c} \dot{\mathbf{R}}_J \cdot \mathbf{A}(t) \right] + \nonumber  \\
	&+& \int_{t_1}^{t_2}\mathrm{d}t \sum_j f_{j\mathbf{k}} w_{\mathbf{k}}
	\left\langle \psi_{j\mathbf{k}}(t) \left|
	\mathrm{i}\frac{\partial}{\partial t}-\hat{H}(\mathbf{r},\mathbf{R},t)
	\right| \psi_{j\mathbf{k}}(t) \right\rangle 	.
	\end{eqnarray}
	$M_J$ is the mass of a nucleus with index $J$, $Z_J$ its atomic number, and $\mathbf{R}_J(t)$ its position. $| \psi_{j\mathbf{k}}(t)\rangle$ is the single-particle KS wavefunction with wavevector $\mathbf{k}$, state index $j$, and occupation $f_{j\mathbf{k}}$. $w_{\mathbf{k}}$ is the weight of the \kpt{.} The time-dependent KS hamiltonian is
	\begin{equation}
	\hat{H}(\mathbf{r},\mathbf{R},t) = \frac{1}{2}\left(-\mathrm{i}\nabla + \frac{1}{c}\mathbf{A}(t)\right)^2
	+v_{H}(\mathbf{r},t)+v_{XC}(\mathbf{r},t) + v_{nucl}(\mathbf{r},\mathbf{R}),
	\end{equation}
	where $v_H$ and $v_{XC}$ are the time-dependent Hartree and exchange-correlation potentials. $v_{nucl}(\mathbf{r},\mathbf{R})$ is the electrostatic potential that comprises all the nuclei-nuclei and nuclei-electron interactions
	\begin{equation}
	v_{nucl}(\mathbf{r},\mathbf{R}) = \displaystyle{ \frac{1}{2}\sum_{I,J}^{I\not=J} \frac{Z_I Z_J}{|\mathbf{R}_I-\mathbf{R}_J|}-\sum_J \frac{Z_J}{|\mathbf{r}-\mathbf{R}_J|} }.
	\end{equation}

	\subsection{Electronic equation of motion}
	\exciting{} employs the LAPW+lo basis \cite{Gulans_2014,Singh}. Around the nuclei, non-overlapping spheres (termed {\it muffin-tin spheres} or {\it atomic spheres}) are defined. The space outside the muffin-tin (MT) spheres is called the {\it interstitial region}. An LAPW function $| \phi_{\mathbf{G}+\mathbf{k}}\rangle$ with wavevector $\mathbf{G}+\mathbf{k}$, where $\mathbf{G}$ is a reciprocal lattice vector and $\mathbf{k}$ a crystal momentum from the first Brillouin zone (BZ), are described by planewaves in the interstitial region, each of them being smoothly augmented into atomic-like wavefunctions inside the atomic spheres. lo's, $| \phi_{\nu}\rangle$, here indexed by $\nu$, are functions defined only inside one muffin-tin sphere and zero outside. In Ehrenfest dynamics, the basis is implicitly time-dependent, since the muffin-tin spheres are centered around the nuclei whose positions evolve in time. The KS wavefunctions $| \psi_{j\mathbf{k}}(t)\rangle$ are expanded in terms of the basis as

	\begin{equation}\label{eq-KSwavefunction-LAPW-lo}
	| \psi_{j\mathbf{k}}(t)\rangle
	=
	\sum_{\mathbf{G}}C_{j\mathbf{k}\mathbf{G}}(t)
	| \phi_{\mathbf{G}+\mathbf{k}}\rangle+
	\sum_\nu C_{j\mathbf{k}\nu}(t)
	| \phi_{\nu}\rangle = \sum_{\mu}C_{j\mathbf{k}\mu}(t) | \phi_{\mu\mathbf{k}}\rangle.
	\end{equation}
	Using the more compact notation indicated by the last summation in Eq. (\ref{eq-KSwavefunction-LAPW-lo}), the time-evolution of the array $C_{j\mathbf{k}}(t)$, obtained by minimizing the integral of action, follows
	\begin{equation}\label{eq-evolution-coefficients}
	S_{\mathbf{k}}(t)\dot{C}_{j\mathbf{k}}(t)	=
	-\mathrm{i}
	H_{\mathbf{k}}(t)
	C_{j\mathbf{k}}(t)
	-
	B_{\mathbf{k}}(t)
	C_{j\mathbf{k}}(t),
	\end{equation}
	where $[S_{\mathbf{k}}(t)]_{\mu\mu'}= \langle \phi_{\mu\mathbf{k}} | \phi_{\mu'\mathbf{k}}\rangle$ and $[H_{\mathbf{k}}(t)]_{\mu\mu'}= \langle \phi_{\mu\mathbf{k}} | \hat{H}(t)| \phi_{\mu'\mathbf{k}}\rangle$ are the overlap and ha\-mil\-tonian matrices, respectively, and $B_{\mathbf{k}}(t)$ reflects the time evolution of the basis:
	\begin{equation}\label{eq-evolution_of_basis}
	B_{\mathbf{k}}(t)
	=
	\sum_J
	\dot{\mathbf{R}}_J\cdot
	\mathcal{B}_{J\mathbf{k}}(t),
	\qquad
	[\mathcal{B}_{J\mathbf{k}}(t)]_{\mu'\mu}
	=
	\left \langle
	\phi_{\mu'\mathbf{k}}\left|
	\frac{\partial \phi_{\mu\mathbf{k}}}{\partial \mathbf{R}_J}\right.
	\right\rangle.
	\end{equation}

	The auxiliary matrix $\mathcal{B}_{J\mathbf{k}}$ is obtained by considering that the changes in the LAPW terms due to the movement of the nuclei are \cite{Yu_1991}:
	\begin{equation}\label{eq-gradient_lapw}
	\left|
	\frac{\partial \phi_{\mathbf{k}+\mathbf{G}}}{\partial \mathbf{R}_J}
	\right\rangle=
	\mathrm{i}(\mathbf{k}+\mathbf{G})
	\left|
	\phi_{\mathbf{k}+\mathbf{G}}
	\right\rangle_{MT_J}-
	\left|
	\nabla
	\phi_{\mathbf{k}+\mathbf{G}}
	\right\rangle_{MT_J},
	\end{equation}
	where the subscript $MT_J$ indicates that the corresponding function is non-zero only inside the MT sphere centered around the nucleus $J$. The analogous expression for the lo's reads \cite{Yu_1991}:
	\begin{equation}\label{eq-gradient_lo}
	\left|
	\frac{\partial \phi_{\nu}}{\partial \mathbf{R}_J}
	\right\rangle=
	-
	\left|
	\nabla
	\phi_{\nu}
	\right\rangle_{MT_J}.
	\end{equation}
	Inserting Eqs. (\ref{eq-gradient_lapw}) and (\ref{eq-gradient_lo}) into Eq. (\ref{eq-evolution_of_basis}), and considering that $\mu$ and $\mu'$ can be LAPWs or lo's, leads to:
	\begin{equation}\label{eq-mathcalB}
	[\mathcal{B}_{J\mathbf{k}}(t)]_{\mu'\mu}=\left\{
	\begin{array}{l}
	\phantom{-}\mathrm{i}(\mathbf{G}+\mathbf{k})
	\langle
	\phi_{\mathbf{G}'+\mathbf{k}}
	|\phi_{\mathbf{G}+\mathbf{k}}
	\rangle_{MT_J}
	-\mathrm{i}
	\langle
	\phi_{\mathbf{G}'+\mathbf{k}}|\mathbf{p}
	|\phi_{\mathbf{G}+\mathbf{k}}
	\rangle_{MT_J}
	\\
	-\mathrm{i}\langle
	\phi_{\mathbf{G}'+\mathbf{k}}|\mathbf{p}
	|\phi_{\mathbf{G}+\mathbf\nu}\rangle_{MT_J}
	\\
	\phantom{-}\mathrm{i}(\mathbf{G}+\mathbf{k})
	\langle
	\phi_{\nu'}
	|\phi_{\mathbf{G}+\mathbf{k}}
	\rangle_{MT_J}-
	\mathrm{i}
	\langle
	\phi_{\nu'}|\mathbf{p}
	|\phi_{\mathbf{G}+\mathbf{k}}
	\rangle_{MT_J}
	\\
	-\mathrm{i}
	\langle
	\phi_{\nu'}|\mathbf{p}
	|\phi_{\nu}
	\rangle_{MT_J}
	\end{array}
	\right.
	\end{equation}
	where $\mathbf{p} = -\mathrm{i}\nabla$ is the momentum operator. The inner products with subscripts $MT_J$ indicate that the integrals which they contain are evaluated only inside $MT_J$.

	\subsection{Equation of nuclear motion}
	Extremizing Eq. (\ref{eq-action}) with respect to $\mathbf{R}_J$ leads to the following equation of motion for the nuclei:
	\begin{equation}\label{eq-motion-nuclei}
	M_J\ddot{\mathbf{R}}_J(t)=\mathbf{F}_{ext,J}(\mathbf{R}_J,t)
	+
	\mathbf{F}_{HF,J}(\mathbf{R}_J,t)
	+
	\mathbf{F}_{corr,J}(\mathbf{R}_J,t).
	\end{equation}
	The external force $\mathbf{F}_{ext,J}$ arises from the interaction with the applied electric field:
	\begin{equation}
	\mathbf{F}_{ext,J}(\mathbf{R}_J,t) = Z_J\mathbf{E}(t)
	=
	-\frac{Z_J}{c}\frac{d \mathbf{A}}{dt}(t).
	\end{equation}
	The Hellmann-Feynman force, $\mathbf{F}_{HF,J}$, can be expressed as \cite{Kohler_1996}
	\begin{eqnarray}
	\mathbf{F}_{HF,J}(\mathbf{R}_J,t) &=&
	-\int \mathrm{d}\mathbf{r}\,
	n(\mathbf{r},t)
	\frac{\partial v_{nucl}}{\partial \mathbf{R}_J}
	\nonumber \\
	&=&
	Z_J\lim_{r_J \to 0}
	\frac{1}{r_J}\sum_{m=-1}^{1}
	\nabla [r_J v_{C,lm}(r_J) Y_{lm}(\hat{r}_J)]_{l=1},
	\end{eqnarray}
	where $\mathbf{r}_J = \mathbf{r}-\mathbf{R}_J$, $v_{C,lm}(r_J)$ is the radial function that is multiplied by $Y_{lm}$ when $v_C = v_{nucl} + v_{H}$ is expanded in spherical harmonics, and the gradient above is taken with respect to $\mathbf{r}_J$.

While the Hellmann-Feynman force is common to implementations of any basis-set type, so-called Pulay corrections need to be considered in the LAPW+lo formalism \cite{Yu_1991,Kohler_1996}. The additional term, $\mathbf{F}_{corr,J}$, arising due to the basis incompleteness, is a sum of contributions from the valence ($\mathbf{F}_{val,J}$) and core ($\mathbf{F}_{core,J}$) states. Neglecting second-order terms with respect to the displacement, one arrives at:
	\begin{eqnarray}
	\mathbf{F}_{core,J}(\mathbf{R}_J,t) &=&
	-\int_{MT_J} \mathrm{d}\mathbf{r}\,
	n_{core}(\mathbf{r},t)
	\nabla v_{KS}(\mathbf{r},t)\nonumber \\
	&=&
	\int_{MT_J} \mathrm{d}\mathbf{r}\,
	v_{KS}(\mathbf{r},t)
	\nabla n_{core}(\mathbf{r},t),
	\end{eqnarray}
	\begin{eqnarray}
	\mathbf{F}_{val,J}(\mathbf{R}_J,t)&=&
	-\sum_{j\mathbf{k}}w_{\mathbf{k}}f_{j\mathbf{k}}
	(C_{j\mathbf{k}})^\dagger
	(
	\mathcal{H}_{J\mathbf{k}}
	-
	\mathcal{S}_{J\mathbf{k}}
	)
	C_{j\mathbf{k}}\nonumber \\
	&+&
	\int_{MT_J}\mathrm{d}\mathbf{r}
	(\nabla n_{val}(\mathbf{r}))\left( v_{KS}(\mathbf{r})
	+\frac{\mathbf{A}^2}{2c^2}
	\right) , \label{eq-Fval}
	\end{eqnarray}
	$n_{core}$ and $n_{val}$ are the density of core and valence electrons, respectively. The auxiliary matrices $\mathcal{S}_{J\mathbf{k}}$ and $\mathcal{H}_{J\mathbf{k}}$ are defined as:
	\begin{equation} \mathcal{S}_{J\mathbf{k}}(t)=
	H_{\mathbf{k}}(t)
	S_{\mathbf{k}}^{-1}(t)
	\mathcal{B}_{J\mathbf{k}}(t)
	+
	\mathcal{B}_{J\mathbf{k}}^{\dagger}(t)
	S_{\mathbf{k}}^{-1}(t)
	H_{\mathbf{k}}(t),
	\end{equation}
	\begin{equation}
	[\mathcal{H}_{J\mathbf{k}}(t)]_{\mu'\mu}
	=
	\left\{
	\begin{array}{l}
	\phantom{-}\mathrm{i}(\mathbf{G}-\mathbf{G}')
	\left[
	\langle \phi_{\mathbf{G}'+\mathbf{k}}|\hat{H}(t)| \phi_{\mathbf{G}+\mathbf{k}} \rangle_{MT_J}
	-
	\lambda_J
	\right]
	\\
	-
	\mathrm{i}(\mathbf{G}'+\mathbf{k})
	\langle \phi_{\mathbf{G}'+\mathbf{k}} |\hat{H}(t)| \phi_{\nu} \rangle_{MT_J}
	\\
	\phantom{-}\mathrm{i}(\mathbf{G}+\mathbf{k})
	\langle \phi_{\nu'}|\hat{H}(t)| \phi_{\mathbf{G}+\mathbf{k}} \rangle_{MT_J}
	\\
	\phantom{-}0 \mbox{ for the case lo-lo}
	\end{array}
	\right.
	\end{equation}
	The definition of the auxiliary function $\lambda_J$ is
	\begin{equation}
	\lambda_J =
	\left(
	\frac{\mathbf{G}'+\mathbf{k}}{2}
	+
	\frac{\mathbf{A}(t)}{c}
	\right)
	\cdot
	(\mathbf{G}+\mathbf{k})
	\frac{4\pi R_{MT_J}^3}{\Omega} \frac{ j_1(|\mathbf{G}-\mathbf{G}'|R_{MT_J})}{|\mathbf{G}-\mathbf{G}'|R_{MT_J}}
	\mathrm{e}^{\mathrm{i}
		\mathbf{(\mathbf{G}-\mathbf{G}')}\cdot \mathbf{R}_J}
	\end{equation}
	where $\Omega$ is the unit cell volume.

	\subsection{Different time steps}
	As suggested in Refs. \cite{Wang_2011} and \cite{Li_2005}, we allow for two different time steps. A smaller one, $\Delta t$, is taken to solve Eq. (\ref{eq-evolution-coefficients}), which dictates the dynamics of the electrons. The positions of the nuclei are kept fixed and are only changed in periods of $\alpha \Delta t$, where $\alpha$ is a positive integer. This is physically grounded on the larger masses of the nuclei, when compared to electrons. At the same time, this procedure permits faster calculations, since the evaluation of the forces on the ions, specially the correction $\mathbf{F}_{val,J}(\mathbf{R}_J,t)$, can be slow. In Section \ref{sec-results}, we will address the convergence behavior with respect to $\alpha$.

	\section{Examples of applications}\label{sec-results}
	\subsection{BN in a distorted lattice}
	Our first example is BN in zinc-blend geometry with the experimental lattice constant of $a=6.833$ bohr \cite{Knittle_1989}. An $8\times 8\times 8$ \kgrid{} and a basis-set cutoff of \rgkmax{=}6.0 are employed in the calculation. As initial condition for the MD, nitrogen is displaced from $(0.25,0.25,0.25)$ to $(0.26,0.26,0.26)$ (in lattice coordinates), while boron is kept at the origin of the unit cell. This gives rise to a restoring force, making the atoms oscillate around their equilibrium positions. For the MD, we employ a time step of 0.01 a.u. for electrons and 0.05~a.u. for the nuclei. The approximated enforced time-reversal symmetry (AETRS) method together with a Taylor expansion up to the 4th order \cite{Pela_2021} are used as propagator for solving Eq. (\ref{eq-evolution-coefficients}).

	\begin{figure}[htb]
	\centering
		\includegraphics[scale=0.8]{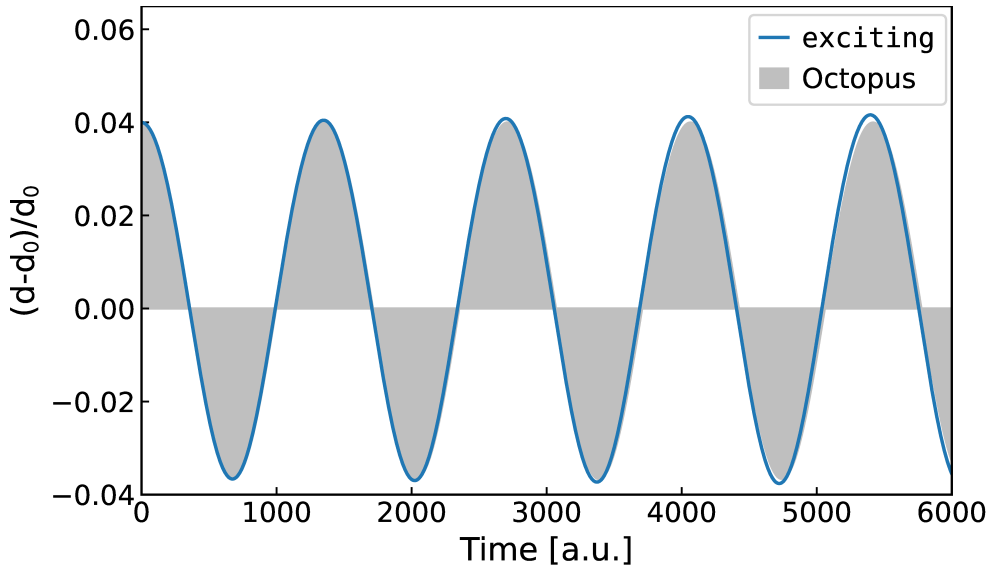}
		\caption{Oscillations of the distance $d$ between boron and nitrogen with respect to the equilibrium bond length $d_0$ -- comparison between \exciting{} and Octopus.}\label{fig-comparison-octopus-BN}
	\end{figure}
	In Fig. \ref{fig-comparison-octopus-BN}, we depict in blue the evolution of the distance $d$ between boron and nitrogen with respect to the equilibrium bond length $d_0$. In gray are the results obtained with Octopus \cite{Tancogne-Dejean_2020} employing a spacing of $a/40$ for the real-space grid. There is very good agreement between the two codes, and the difference in the period of oscillations, provided in Table \ref{tab-period}, is less than $0.37$\%.
	\begin{table}[htb]
		\centering
		\caption{Period $T$ (in a.u.) of oscillations in cubic BN (Fig. \ref{fig-comparison-octopus-BN}) and diamond (Fig. \ref{fig-comparison-octopus-C}) obtained by MD. GS indicates the expected period from a separate set of ground-state calculations (see text). }\label{tab-period}
		\begin{tabular}{ccc}
			\\ \hline
			&\multicolumn{2}{c}{$T$ [a.u.]}\\
			& BN & C \\  \hline
			\exciting{} - MD & 1349 & 1239\\
			Octopus & 1354 & 1242\\
			\exciting{} - GS & 1294 & 1091\\ \hline
		\end{tabular}
	\end{table}
	In this Table, we also provide the period predicted through $T=2\pi\sqrt{\mu/k}$, where $\mu$ is the reduced mass of boron and nitrogen, and $k$ is the spring constant obtained from a set of separate ground-state (GS) calculations in which B and N are displaced from each other, and the total energy with respect to the displacement is fitted with a parabola. The period predicted from these GS calculations differs by approximately 4.0\% from the other two results. We attribute this difference to non-linear effects. Even though the initial condition of placing N at $(0.26,0.26,0.26)$ may seem a small perturbation, it is enough to reach the non-linear regime. This can be seen in Fig. \ref{fig-comparison-octopus-BN}, since the oscillations are not symmetric with respect to the equilibrium position.

    \subsection{Diamond in a distorted lattice}\label{subsec-diamond}
	As a second example, we consider diamond with a distortion as initial condition for the MD, similar to the previous subsection. We employ the experimental lattice constant $a=6.7407$ bohr \cite{Hom_1975}. Like in the example of BN, we displace one carbon atom to a non-equilibrium position, in this case $(0.30,0.30,0.30)$, while keeping the other C atom at the origin. The restoring force is expected to be larger than is the previous case and to drive further in the non-linear regime. At first, we use the same time step of 0.05 a.u. for electrons and nuclei. The AETRS method together with a Taylor expansion up to the 4th order \cite{Pela_2021} are used as propagator for solving Eq. (\ref{eq-evolution-coefficients}).
	\begin{figure}[htb]
		\includegraphics[scale=1]{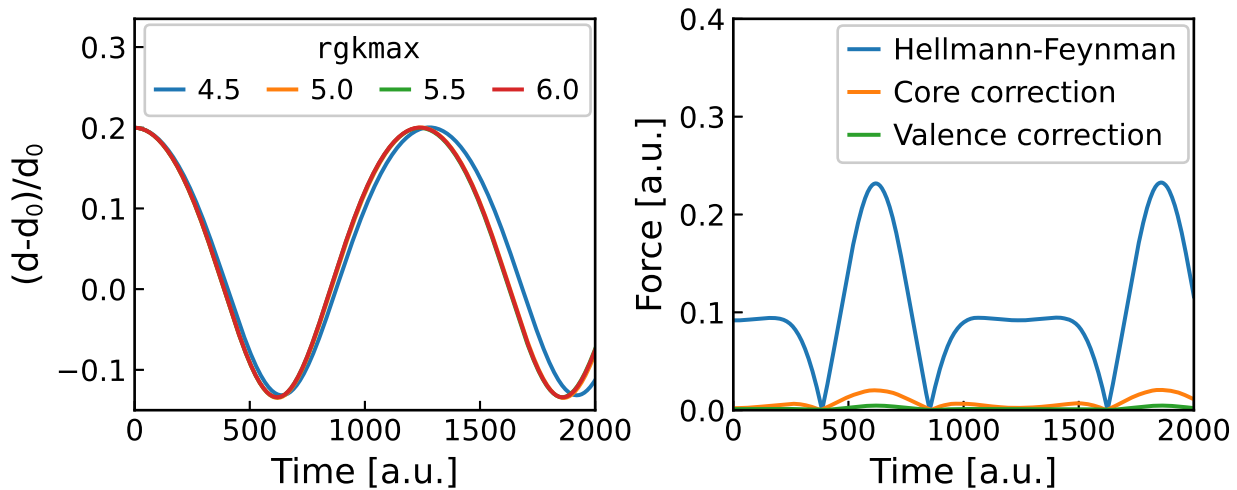}
		\caption{Left: Impact of the basis set size, measured by the parameter \rgkmax{}, on the oscillations of the carbon-carbon distance $d$ relative to the equilibrium bond length $d_0$. Right: Absolute value of the contributions to the force acting on the displaced carbon atom over time (\rgkmax{=}6.0).}
		\label{fig-C-rgkmax}
	\end{figure}
	Our calculations employ an $8\times 8\times 8$ \kgrid{}, and \rgkmax{} is varied from 4.5 to 6.0, as shown in Fig. \ref{fig-C-rgkmax}.
	We observe that starting from an \rgkmax{} of 5.0, the curves are almost indistinguishable. On the right side of Fig. \ref{fig-C-rgkmax}, for the case \rgkmax{=}6.0, we compare the absolute value of the forces given in Eq. (\ref{eq-motion-nuclei}): $\mathbf{F}_{HF,J}(\mathbf{R}_J,t)$ and the correction $\mathbf{F}_{corr,J}(\mathbf{R}_J,t)$ split into $\mathbf{F}_{core,J}(\mathbf{R}_J,t)$ and $\mathbf{F}_{val,J}(\mathbf{R}_J,t)$. We see that the Hellmann-Feynman term is much larger than the corrections. In this case, \rgkmax{=}6.0 makes the LAPW basis largely complete.

\begin{figure}[htb]
		\includegraphics[scale=1]{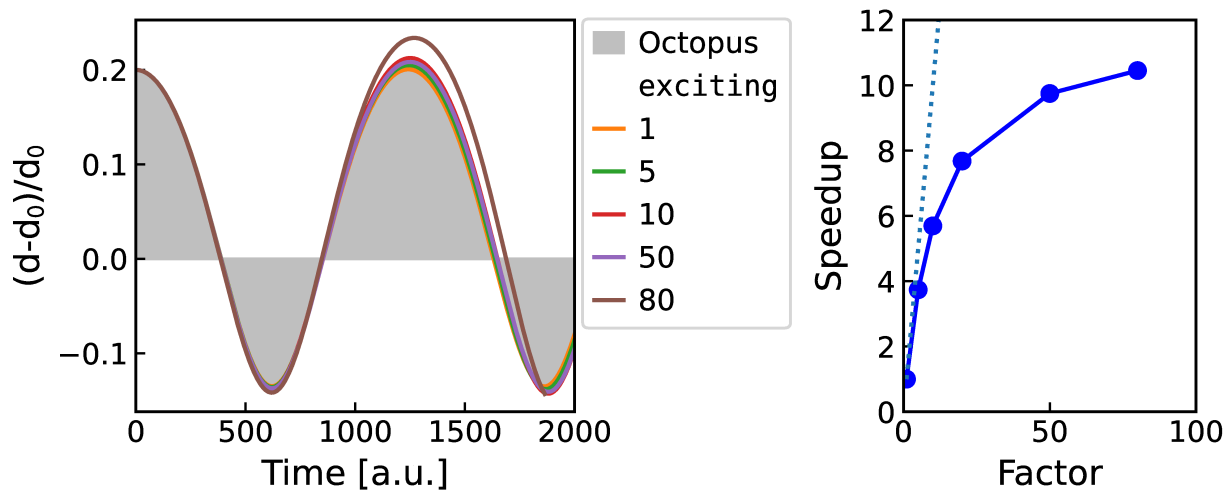}
		\caption{Left: Influence of the multiplicative factor $\alpha$ on the MD of diamond, if we update the atomic positions with a time step $\alpha \Delta t$, $\Delta t$ being the time step for the electronic dynamics within RT-TDDFT. Right: Speedup as a function of $\alpha$; the dotted line indicates the ideal speedup. }\label{fig-comparison-octopus-C}
	\end{figure}

	Figure \ref{fig-comparison-octopus-C} displays on the left the impact of the time-step multiplier $\alpha$ on the oscillations of the carbon atoms, when we consider time steps $\alpha \Delta t$ and $\Delta t$, respectively,  for updating the nuclei positions and the KS wavefunctions. On the right, the speedup as a function of $\alpha$ is shown. It is apparent that factors of 5 or 10 are possible without losing significant precision.
	A calculation performed with Octopus using a real-space grid with spacing of $a/25$ is shown on the left side of Fig. \ref{fig-comparison-octopus-C} for comparison and highlighting remarkable agreement for $\alpha=1$. In Table \ref{tab-period}, we show the period obtained with \exciting{} (same time step for nuclei and electrons) and Octopus, the difference being less than $0.25$\%. Considering, as in the case of BN, a set of GS calculations to predict the period from $T=2\pi\sqrt{\mu/k}$, we observe a higher discrepancy  of 12\%, owing to larger non-linear effects that are also seen in the asymmetry of the oscillations around the equilibrium position (Figs. \ref{fig-C-rgkmax} and \ref{fig-comparison-octopus-C}).

	\subsection{BN exposed to an intense electric field}
	In our third example, we start from the groundstate of zinc-blend BN and apply an electric field $\mathbf{E}(t) = -\frac{1}{c} \frac{d \mathbf{A}}{ dt }$ along the [111] direction, where $\mathbf{A}(t) = \mathbf{A}_0 f(t) \cos(\omega_0 t).$
	To be able to observe a non-negligible effect of the electric field on the ions, we choose an angular frequency $\omega_0$=0.5 a.u and set the magnitude of $\mathbf{A}_0$ to 10.0~a.u. in each cartesian direction. For the envelope function $f(t)$, we consider $\sin^2( \pi t/T_p)$ until a time $T_p=1000$~a.u., and zero afterwards. This function mimics a gaussian envelope, quite common in experiments.

	\begin{figure}[htb]
	  \includegraphics[scale=0.9]{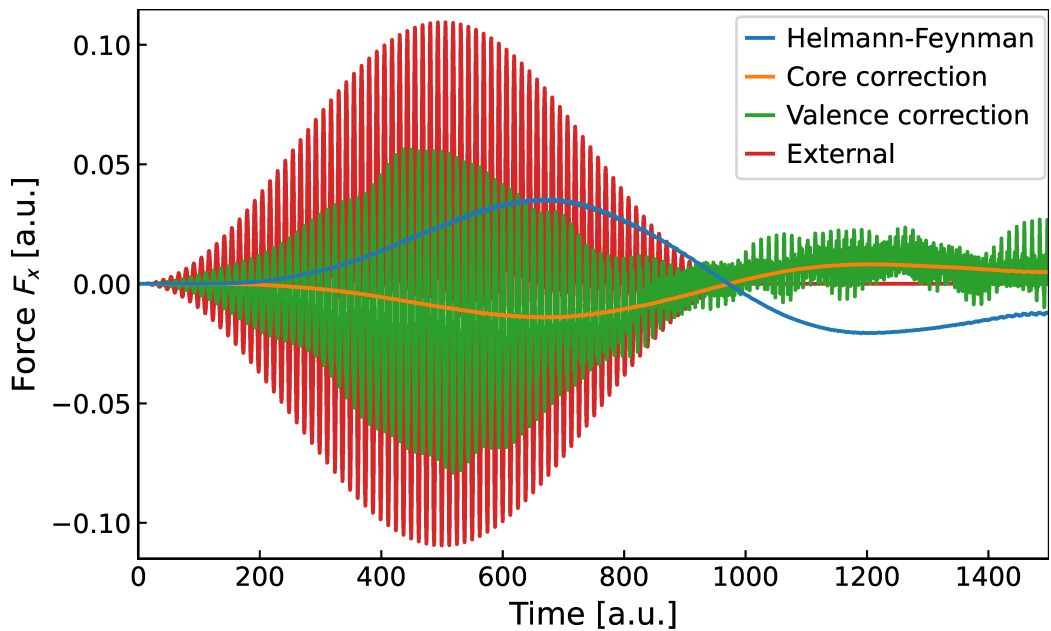}
	  \caption{Different contributions to the force felt by the boron atom in BN. Only the $x$ component is shown, since symmetry implies that the other components are equivalent. }\label{fig-BN-MD-Laser-Forces}
	\end{figure}

	In Fig. \ref{fig-BN-MD-Laser-Forces}, we display the $x$ component of the different contributions to the force felt by boron as given in Eqs. (\ref{eq-motion-nuclei})-(\ref{eq-Fval}): $F_{HF}$, $F_{core}$, $F_{val}$, and $F_{ext}$. The amplitudes of all contributions are small compared to the example in Section \ref{subsec-diamond} (see Fig. \ref{fig-C-rgkmax}), where the diamond lattice is artificially heavily distorted, a situation difficult to be reproduced by applying a laser. The external force $F_{ext}$ assumes, as expected, the shape of the applied field. Rapid oscillations are also observable in $F_{val}$ that explicitly depends on the vector potential $\mathbf{A}(t)$. These oscillations still remain after the field is no longer acting on the system, as a result of $\mathbf{A}_0$ being high enough to excite to the non-linear regime. $F_{HF}$ varies smoothly, since the interaction with the other nuclei is expected to follow the timescale of the phonon modes, which is on the order of 1000~a.u. \cite{Krystian_1997}. The same applies to $F_{core}$: Since we disregard the effect of $\mathbf{A}(t)$ on the core electrons, $F_{core}$ should arise mainly from displacements of the nuclei, which carry the core electrons along.

\begin{figure}[tbb]
		\includegraphics[scale=0.9]{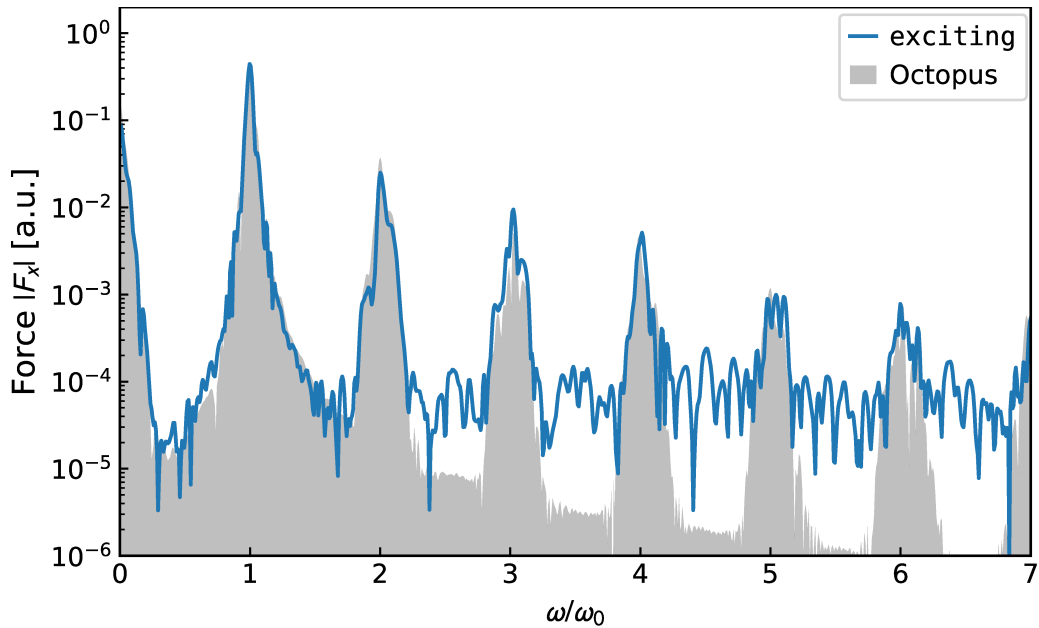}
		\caption{Comparison of \exciting{} and Octopus: Fourier transform of the force acting on boron (for $\omega_0$=0.5 a.u.).}\label{fig-BN-MD-Laser-comparison-Octopus}
	\end{figure}

In Fig. \ref{fig-BN-MD-Laser-comparison-Octopus}, we compare the force on boron obtained with \exciting{} and  Octopus. Since the force oscillates rapidly as shown in Fig. \ref{fig-BN-MD-Laser-Forces}, we display its Fourier transform for better comparison. Overall, we find good agreement.

	\section{Conclusions}\label{sec-conclusions}
	In this paper, we have reported the implementation of Ehrenfest dynamics combined with RT-TDDFT in the all-electron full-potential package \exciting{} that implements the LAPW+lo basis. To illustrate its performance, we have considered three examples: diamond and cubic BN with an initial lattice distortion, and cubic BN excited by an oscillating electric field. Comparison of our results with the Octopus code shows good agreement. The code will be freely available in the next release of \exciting{} (fluorine), and all results underlying this publication are available on the NOMAD Laboratory  \cite{Draxl_2019,Draxl_2018,nomad-doi}.

		\ack
		This work was supported by the German Research Foundation within project Nr. 182087777 - SFB 951 (B11) and the priority program SPP2196 “Perovskite Semiconductors,” project Nr. 424709454.

	\section*{References}
	\bibliographystyle{iopart-num}
	\bibliography{references}

\providecommand{\newblock}{}
\begin{thebibliography}{10}
\expandafter\ifx\csname url\endcsname\relax
  \def\url#1{{\tt #1}}\fi
\expandafter\ifx\csname urlprefix\endcsname\relax\def\urlprefix{URL }\fi
\providecommand{\eprint}[2][]{\url{#2}}

\bibitem{Alder_1959}
Alder B~J and Wainwright T~E 1959 {\em The Journal of Chemical Physics\/} {\bf
  31} 459--466

\bibitem{Andrade_2009}
Andrade X, Castro A, Zueco D, Alonso J~L, Echenique P, Falceto F and Rubio A
  2009 {\em Journal of Chemical Theory and Computation\/} {\bf 5} 728--742

\bibitem{Scherrer_2017}
Scherrer A, Agostini F, Sebastiani D, Gross E and Vuilleumier R 2017 {\em
  Physical Review X\/} {\bf 7} 031035 ISSN 2160-3308

\bibitem{Nelson_2020}
Nelson T~R, White A~J, Bjorgaard J~A, Sifain A~E, Zhang Y, Nebgen B,
  Fernandez-Alberti S, Mozyrsky D, Roitberg A~E and Tretiak S 2020 {\em
  Chemical Reviews\/} {\bf 120} 2215--2287

\bibitem{Curchod_2013}
Curchod B~F~E, Rothlisberger U and Tavernelli I 2013 {\em ChemPhysChem\/} {\bf
  14} 1314--1340 ISSN 1439-7641

\bibitem{Otero_2018}
Crespo-Otero R and Barbatti M 2018 {\em Chemical Reviews\/} {\bf 118}
  7026--7068 ISSN 0009-2665

\bibitem{Pradhan_2018}
Pradhan E, Sato K and Akimov A~V 2018 {\em Journal of Physics: Condensed
  Matter\/} {\bf 30} 484002 ISSN 0953-8984

\bibitem{You_2021}
You P, Chen D, Lian C, Zhang C and Meng S 2021 {\em Wiley Interdisciplinary
  Reviews: Computational Molecular Science\/} {\bf 11} ISSN 1759-0876

\bibitem{Ojanpera_2012}
Ojanperä A, Havu V, Lehtovaara L and Puska M 2012 {\em The Journal of Chemical
  Physics\/} {\bf 136} 144103

\bibitem{Wang_2011}
Wang F, Yam C~Y, Hu L and Chen G 2011 {\em The Journal of Chemical Physics\/}
  {\bf 135} 044126

\bibitem{Andermatt_2016}
Andermatt S, Cha J, Schiffmann F and VandeVondele J 2016 {\em Journal of
  Chemical Theory and Computation\/} {\bf 12} 3214

\bibitem{Isborn_2007}
Isborn C~M, Li X and Tully J~C 2007 {\em The Journal of Chemical Physics\/}
  {\bf 126} 134307

\bibitem{Miyamoto_2006}
Miyamoto Y, Rubio A and Tom\'anek D 2006 {\em Phys. Rev. Lett.\/} {\bf 97}(12)
  126104 \urlprefix\url{https://link.aps.org/doi/10.1103/PhysRevLett.97.126104}

\bibitem{Krasheninnikov_2007}
Krasheninnikov A~V, Miyamoto Y and Tom\'anek D 2007 {\em Phys. Rev. Lett.\/}
  {\bf 99}(1) 016104
  \urlprefix\url{https://link.aps.org/doi/10.1103/PhysRevLett.99.016104}

\bibitem{Prezhdo_2009}
Prezhdo O~V 2009 {\em Accounts of Chemical Research\/} {\bf 42} 2005--2016
  pMID: 19888715

\bibitem{Long_2011}
Long R and Prezhdo O~V 2011 {\em Journal of the American Chemical Society\/}
  {\bf 133} 19240--19249 pMID: 22007727

\bibitem{Long_2014}
Long R and Prezhdo O~V 2014 {\em Nano Letters\/} {\bf 14} 3335--3341

\bibitem{Weile_2018}
Jia W, An D, Wang L~W and Lin L 2018 {\em Journal of Chemical Theory and
  Computation\/} {\bf 14} 5645--5652 ISSN 1549-9618

\bibitem{Lejaeghereaad_2016}
Lejaeghere K, Bihlmayer G, Bj{\"o}rkman T, Blaha P, Bl{\"u}gel S, Blum V,
  Caliste D, Castelli I~E, Clark S~J, Dal~Corso A, de~Gironcoli S, Deutsch T,
  Dewhurst J~K, Di~Marco I, Draxl C, Du{\l}ak M, Eriksson O, Flores-Livas J~A,
  Garrity K~F, Genovese L, Giannozzi P, Giantomassi M, Goedecker S, Gonze X,
  Gr{\r a}n{\"a}s O, Gross E~K~U, Gulans A, Gygi F, Hamann D~R, Hasnip P~J,
  Holzwarth N~A~W, Iu{\c s}an D, Jochym D~B, Jollet F, Jones D, Kresse G,
  Koepernik K, K{\"u}{\c c}{\"u}kbenli E, Kvashnin Y~O, Locht I~L~M, Lubeck S,
  Marsman M, Marzari N, Nitzsche U, Nordstr{\"o}m L, Ozaki T, Paulatto L,
  Pickard C~J, Poelmans W, Probert M~I~J, Refson K, Richter M, Rignanese G~M,
  Saha S, Scheffler M, Schlipf M, Schwarz K, Sharma S, Tavazza F, Thunstr{\"o}m
  P, Tkatchenko A, Torrent M, Vanderbilt D, van Setten M~J, Van~Speybroeck V,
  Wills J~M, Yates J~R, Zhang G~X and Cottenier S 2016 {\em Science\/} {\bf
  351} ISSN 0036-8075

\bibitem{Gulans_2018}
Gulans A, Kozhevnikov A and Draxl C 2018 {\em Phys. Rev. B\/} {\bf 97}(16)
  161105

\bibitem{Gulans_2014}
Gulans A, Kontur S, Meisenbichler C, Nabok D, Pavone P, Rigamonti S, Sagmeister
  S, Werner U and Draxl C 2014 {\em Journal of Physics: Condensed Matter\/}
  {\bf 26} 363202

\bibitem{exciting_webpage}
The exciting code \url{http://exciting-code.org/}

\bibitem{Pela_2021}
Pela R~R and Draxl C 2021 {\em Electronic Structure\/} {\bf 3} 037001

\bibitem{Saalmann_1996}
Saalmann U and Schmidt R 1996 {\em Zeitschrift für Physik D Atoms, Molecules
  and Clusters\/} {\bf 38} 153--163

\bibitem{Kunert_2003}
Kunert T and Schmidt R 2003 {\em The European Physical Journal D - Atomic,
  Molecular, Optical and Plasma Physics\/} {\bf 25} 15--24

\bibitem{Singh}
Singh D~J and Nordstr\"{o}m L 2006 {\em Planewaves, Pseudopotentials, and the
  LAPW Method\/} 2nd ed (New York: Springer) ISBN 0-387-28780-9

\bibitem{Yu_1991}
Yu R, Singh D and Krakauer H 1991 {\em Phys. Rev. B\/} {\bf 43}(8) 6411--6422
  \urlprefix\url{https://link.aps.org/doi/10.1103/PhysRevB.43.6411}

\bibitem{Kohler_1996}
Kohler B, Wilke S, Scheffler M, Kouba R and Ambrosch-Draxl C 1996 {\em Computer
  Physics Communications\/} {\bf 94} 31--48 ISSN 0010-4655 (\textit{Preprint}
  \eprint{mtrl-th/9511002})

\bibitem{Li_2005}
Li X, Tully J~C, Schlegel H~B and Frisch M~J 2005 {\em The Journal of Chemical
  Physics\/} {\bf 123} 084106

\bibitem{Knittle_1989}
Knittle E, Wentzcovitch R~M, Jeanloz R and Cohen M~L 1989 {\em Nature\/} {\bf
  337} 349

\bibitem{Tancogne-Dejean_2020}
Tancogne-Dejean N, Oliveira M~J~T, Andrade X, Appel H, Borca C~H, Le~Breton G,
  Buchholz F, Castro A, Corni S, Correa A~A, De~Giovannini U, Delgado A, Eich
  F~G, Flick J, Gil G, Gomez A, Helbig N, Hübener H, Jestädt R, Jornet-Somoza
  J, Larsen A~H, Lebedeva I~V, Lüders M, Marques M~A~L, Ohlmann S~T, Pipolo S,
  Rampp M, Rozzi C~A, Strubbe D~A, Sato S~A, Schäfer C, Theophilou I, Welden A
  and Rubio A 2020 {\em The Journal of Chemical Physics\/} {\bf 152} 124119

\bibitem{Hom_1975}
Hom T, Kiszenik W and Post B 1975 {\em Journal of Applied Crystallography\/}
  {\bf 8} 457--458

\bibitem{Krystian_1997}
Karch K and Bechstedt F 1997 {\em Physical Review B\/} {\bf 56} 7404--7415 ISSN
  1098-0121

\bibitem{Draxl_2019}
Draxl C and Scheffler M 2019 {\em Journal of Physics: Materials\/} {\bf 2}
  036001 \urlprefix\url{https://doi.org/10.1088/2515-7639/ab13bb}

\bibitem{Draxl_2018}
Draxl C and Scheffler M 2018 {\em MRS Bulletin\/} {\bf 43} 676–682

\bibitem{nomad-doi}
Nomad repository, dataset: Md-exciting-octopus.
  \url{https://dx.doi.org/10.17172/NOMAD/2021.12.28-1}

\end{thebibliography}

%	\appendix
%	\section{Expressions}

\end{document}